\documentclass{aa}  

\makeatletter
\renewcommand*\aa@pageof{, page \thepage{} of \pageref*{LastPage}}
\makeatother

\usepackage[utf8]{inputenc}
\usepackage{xcolor}
\usepackage{listings}
\usepackage{graphicx}
\usepackage{txfonts}
\usepackage{hyperref}
\usepackage{paralist}
\defcitealias{Portail:17}{P17}

\newcommand{\ie}{{\it i.e.}\xspace}

\newcommand{\kpc}{\ensuremath{\,{\rm kpc}}\xspace}

\newcommand{\kms}{\ensuremath{\,{\rm km}\,{\rm s}^{-1}}\xspace}
\newcommand{\logg}{\ensuremath{\log (g)}\xspace}
\newcommand{\Teff}{\ensuremath{T_{\rm eff}}\xspace}
\newcommand{\Teffref}{\ensuremath{T_{\rm eff}^{\rm ref}}\xspace}

\newcommand{\K}{\ensuremath{{K_s}}\xspace}

\newcommand{\FeH}{\ensuremath{[{\rm Fe/H}]}\xspace}

\newcommand\figurefraction{1}

\begin{document}

   \title{The chemistry of stars in the bar of the Milky Way}
   \titlerunning{Chemistry of stars in the bar}
   \authorrunning{Wegg, Rojas-Arriagada, Schultheis, Gerhard}
   \subtitle{}

   \author{C. Wegg\inst{1,2}, A. Rojas-Arriagada\inst{3,4}, M. Schultheis\inst{1}
          \and
          O. Gerhard\inst{2}
          }

   \institute{Universit\'e C\^ote d'Azur, Observatoire de la C\^ote d'Azur, CNRS, Laboratoire Lagrange, France\\
              \email{chris.wegg@oca.eu; chriswegg@gmail.com}
         \and
             Max-Planck-Institut f{\"u}r Extraterrestrische Physik, Gie{\ss}enbachstra{\ss}e, D-85748 Garching, Germany
             \and
             Instituto de Astrof\'{i}sica, Facultad de F\'{i}sica, Pontificia Universidad Cat\'olica de Chile, Av. Vicu\~na Mackenna 4860, Santiago, Chile  
\and
Millennium Institute of Astrophysics, Av. Vicu\~{n}a Mackenna 4860, Santiago, Chile
             }

   \date{Received Sep 23, 2019; accepted ...} 

 
  \abstract
  {We use a sample of 938 red clump giant stars located in the direction of the galactic long bar to study the chemistry of Milky Way bar stars. Kinematically separating stars on bar orbits from stars with inner disc orbits, we find that stars on bar-like orbits are more metal rich with a mean iron abundance of $\langle \FeH \rangle = +0.30$ compared to $\langle \FeH \rangle = +0.03$ for the inner disc. Spatially selecting bar stars is complicated by a strong vertical metallicity gradient of $-1.1{\rm dex}/\kpc$, but we find the metallicity distribution varies in a manner consistent with our orbital selection. Our results have two possible interpretations.  The first is that the most metal rich stars in the inner Galaxy pre-existed the bar, but were kinematically cold at the time of bar formation and therefore more easily captured onto bar orbits when the bar formed. The second is that the most metal rich stars formed after the bar, either directly onto the bar following orbits or were captured by the bar after their formation.}


   \maketitle

\section{Introduction}
The vertically extended bulge region of the Milky Way is shaped as a box/peanut bulge \citep[B/P bulge;][]{LopezCorredoira:05,Saito:11,Wegg:13}. Such shapes naturally arise in simulations of barred galaxies \citep[e.g.][]{Athanassoula:05,Inma:06}, and they are commonly seen in external galaxies \citep{Bureau:06,Laurikainen:14}. In these simulations and in external galaxies the bar becomes thinner outside of the central B/P bulge \citep{Erwin:13}. An equivalent longer, thinner bar structure also exists in the Milky Way outside the bulge; this structure is termed the long bar \citep{Hammersley:94,Hammersley:00}. However, because the long bar has comparable thickness to the surrounding disc, with consequently higher extinction, fundamental parameters such as its length and orientation and even relationship to the barred bulge have been difficult to determine \citep{LopezCorredoira:06,CabreraLavers:08,Wegg:15}.

In this work, we study the chemistry of stars in the long bar of the Milky Way from a spectroscopic sample of $\sim$2500 stars. From these we select $\sim$1000 red clump giants (RCGs) whose standard candle nature provides $\sim10\%$ accurate distances. When combined with Gaia proper motions, this provides a homogeneous sample of long bar stars with full 6D kinematic information. We use this, together with spectroscopic iron abundances, to study the long bar and compare it to the surrounding inner disc. The recent work of \citet{Bovy:19} found, using APO Galactic Evolution Experiment (APOGEE) data on the inner Galaxy, that the bar was more metal poor than the surrounding disc. We study a smaller region of sky, which has\ a more homogeneous sample, and find contrasting results.

We first describe our sample of stars and their spectroscopic analysis in \autoref{sec:sample}. In \autoref{sec:rcgselection} we describe the construction of our sample of RCGs. In \autoref{sec:barchem} we make spatial and kinematic selections of stars that are likely to be bar stars and show the chemistry of these bar stars compared to the inner disc of the Milky Way. In \autoref{sec:disc} we discuss the implications of our results.

\section{Sample of stars in the inner Galaxy}
\label{sec:sample}

\subsection{Target selection}

\begin{figure*}
    \centering
    \includegraphics[width=\textwidth]{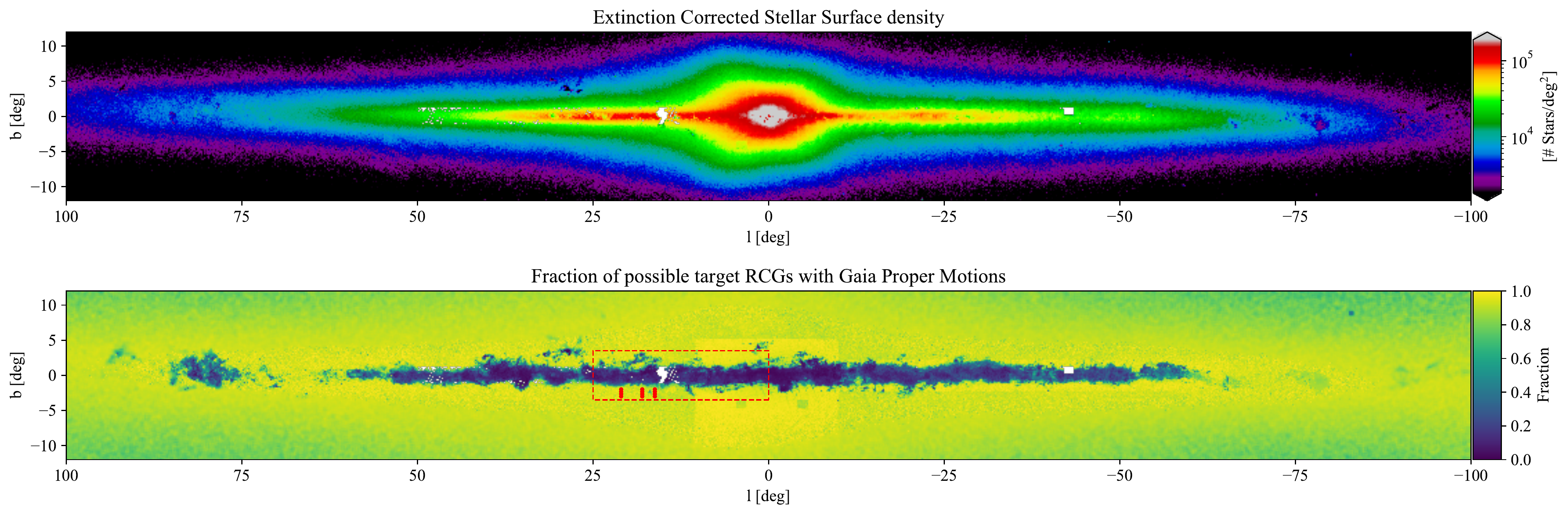}
    \caption[]{{\bf Upper panel:} Surface density of stars with extinction corrected magnitude consistent with being red clump stars 5-8\kpc from the Sun (specifically $13.5<\mu_K<14.5$ in \autoref{eq:mujhk}). Several features are notable: (i) the B/P bulge of the Milky\ Way in the central $10\deg;$ (ii) the higher surface density of stars at $l>0$ compared to $l<0$ outside the bulge near to the galactic plane, indicative of the extension of the bar outside the bulge; and (iii) the tilt of the galactic plane towards the edge of the plot at $|l|>70\deg,  $ which is indicative of the warp/ripples in the outer disc of the Milky Way. We used near-infrared (NIR) data from the VVV\footnotemark survey \citep{Smith:17} where available, falling back to UKIDSS-GPS\footnotemark data \citep{Lucas:08}, and finally 2MASS\footnotemark \citep{Skrutskie:06}. After combining these three datasets only a few small regions lacked NIR of sufficient depth and these are left white. {\bf Lower panel:} Fraction of stars in the upper panel which have proper motions in Gaia DR2. Outside the $\sim 1 \deg$ thick dust layer in the inner Galaxy, almost all possible target RCGs have Gaia proper motions. Our selected fields, shown in red,  are located in the bar outside the bulge. We show a zoom of the red dashed region containing our fields in \autoref{fig:findingchartzoom}. }
    \label{fig:findingchart}
\end{figure*}

\begin{figure}
    \centering
    \includegraphics[width=\figurefraction\columnwidth]{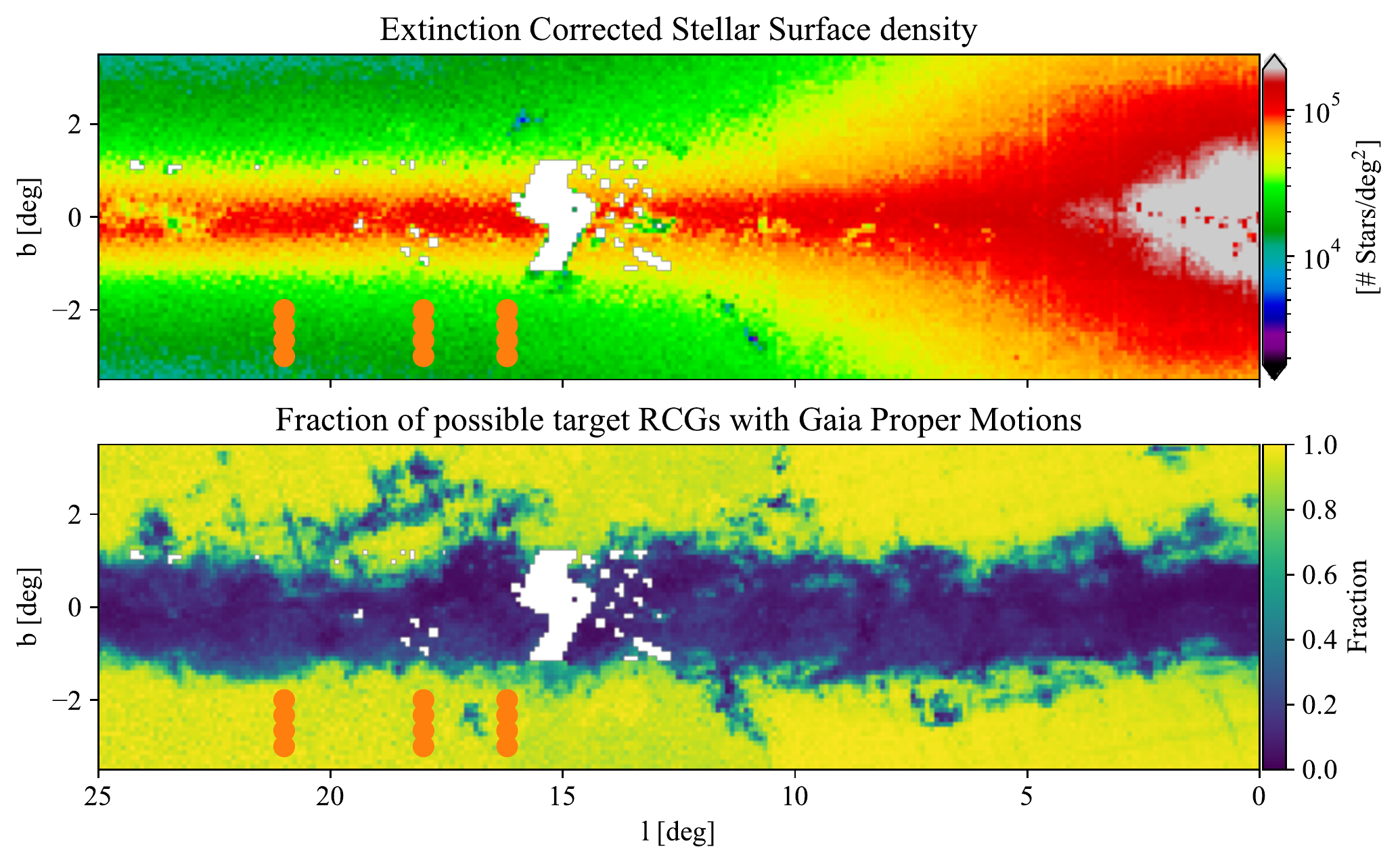}
    \caption{Zoom of \autoref{fig:findingchart} to the region containing our fields. The 12 selected fields, shown in orange, are positioned outside the bulge close to the galactic plane where the bar is still visible in star counts, but where almost all target stars have Gaia proper motions available.}
    \label{fig:findingchartzoom}
\end{figure}

We selected fields in the bar region of the Milky Way, targeting low extinction regions close to the galactic plane, where the bar is most prominent in star counts (see \autoref{fig:findingchart}). Field centres were placed at longitudes of $l=16.2,18,21\deg$, each with latitudes of $b=-2.0,-2.33,-2.66,-3.0\deg$, for a total of 12 fields (see \autoref{fig:findingchartzoom}).

In order to construct a clean sample of bar stars with accurate distances, we targeted standard candle RCGs for which $\sim 10\%$ accurate distances can be calculated \citep{Bovy:14}. In each field we selected stars whose extinction free \K band magnitude would place them in the bar if they were RCGs. Specifically we targeted stars whose $(H-\K)$ extinction free magnitude would place then at $\sim 4-8\kpc$ \citep[i.e. we used eqn 1 of ][]{Wegg:15}. To avoid biasing the sample we did not make a strict colour cut, removing only the small fraction of stars with $(H-\K) < 0.06$. These stars are too blue to be RCGs even without extinction. 
We note that at our chosen latitudes almost all our target RCGs have Gaia astrometry available (Figures \ref{fig:findingchart} and \ref{fig:findingchartzoom}).

\subsection{Spectral analysis}

Spectra were obtained with the Fibre Large
Array Multi Element Spectrograph of the Very Large Telescope at ESO  \citep[ESO/VLT/FLAMES;][]{Pasquini:00} in the Medusa mode of the GIRAFFE multi-object spectrograph. The HR21 set-up was employed, providing a spectral coverage spanning from 8484 to 9001\AA~with a resolving power of $R\sim 18000$. This set-up covers the calcium triplet region and was chosen as it is also a set-up used by the Gaia-ESO Survey (GES) \citep{Gilmore:12} and therefore facilitates comparison with other regions of the Galaxy.

\footnotetext{VISTA (Visible and Infrared Survey Telescope for Astronomy) Variables in the Via Lactea}
\footnotetext{United Kingdom Infrared Deep Sky Survey - Galactic Plane Survey}
\footnotetext{Two Micron All-Sky Survey}

The sample comprises 2456 spectra of unique stars that have a mean signal-to-noise ratio (S/N) of 40. The raw data were reduced through flat-fielding, extraction, wavelength calibration, and correction to heliocentric reference system by ESO as part of the GIRAFFE stream release.

We used the IRAF\footnote{IRAF is distributed by the National Optical Astronomy Observatory, which is operated by the Association of Universities for Research in Astronomy (AURA) under cooperative agreement with the National Science Foundation} task {\it{skytweak}} to remove the sky emission lines from the individual spectra by adopting a median sky spectrum computed from individual spectra of dedicated fibres at each pointing.
After this, we applied the {\it{continuum}} task to normalise the stellar continuum to the unit through a cubic spline fit.
The radial velocity of each individual spectrum was derived by cross correlation against a coarse grid of synthetic templates covering the parameter space of FGK type stars. The radial velocity pipeline selects in a first iteration the best template to the observed spectrum in a $\chi^2$ sense, which is then used to re-compute the final radial velocity.

The set of fundamental parameters ($\Teff$, $\logg$, [M/H]) and global alpha-elements enhancement [$\alpha$/Fe] were obtained by comparing the observed spectra against a grid of synthetic models. We adopted the synthetic spectral library of the GES \citep{deLaverny:12} and the code FERRE. This code performs a global fit search for the set of fundamental parameters providing the best agreement (by minimisation of $\chi^2$) between the observed spectrum and the synthetic grid, allowing for interpolation between the models. During the $\chi^2$ minimisation search, both the models and observed spectra are re-normalised by applying a fourth-degree polynomial. This is intended to be a linear and homogeneous process (i.e. without applying any kind of iterative pixel rejection process), thereby decreasing normalisation systematic differences between observed and model spectra. We tested a couple of different strategies for re-normalisation (using a sample of GES spectra; see below) and we found that our results do not change significantly with the adopted strategy, and are of better quality than those obtained without any re-normalisation. A pixel weight mask was constructed to avoid fitting the core of Ca II triplet (CaT) lines whose modelling suffer from larger uncertainties. A number of other small spectral regions were discarded because of their systematically larger residuals, many of which are attributable to residuals of sky line subtraction.

\begin{figure}
    \centering
    \includegraphics[width=\figurefraction\columnwidth]{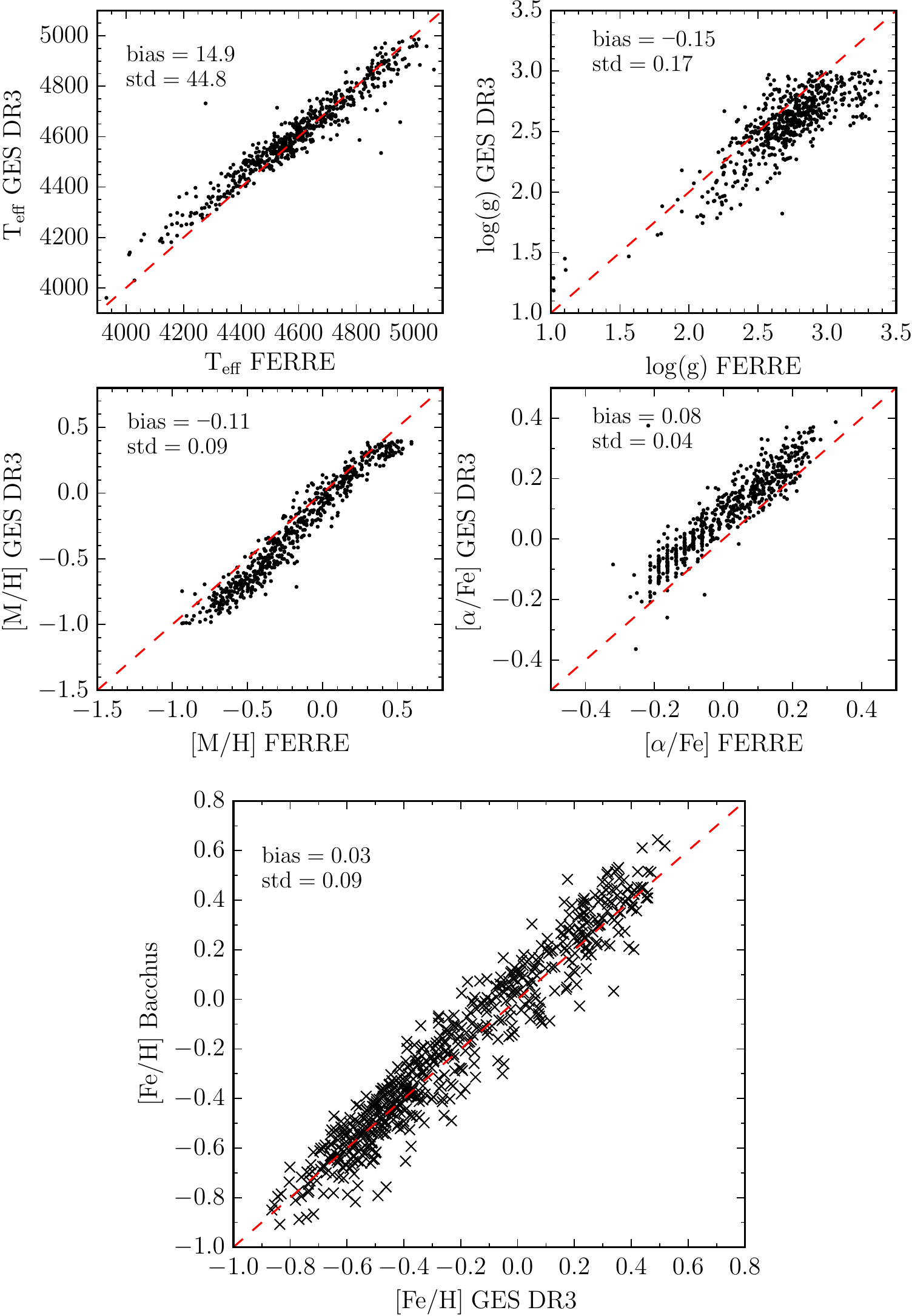}
    \caption{Comparison of fundamental parameters and final iron abundances obtained for a sample of $\sim600$ stars from the GES DR3. In all panels we compare the results obtained by applying to the spectra our spectral analysis procedure to those released as part of the GES DR3. In the four upper panels, we compare results for the global fundamental parameters ${\rm T_{eff}}$, ${\rm \log(g)}$, [M/H], and [$\alpha$/Fe]. In the large lower panel, we compare GES recommended metallicity with those we measured with Bacchus, both determined from iron line-by-line measurements.}
    \label{fig:ferre_ges_comp}
\end{figure}

In order to verify the quality of the fitted stellar parameters and abundances obtained with our procedure, we processed a sample of spectra from the GES in the same way as our dataset.
We considered stars whose set of recommended fundamental parameters (${\rm T_{eff}}$, ${\rm \log(g)}$, [M/H], [$\alpha$/Fe]), and metallicity [Fe/H] were available from the third public data release (DR3). From this, we selected a random sample of $\sim600$ bulge giant ($\log(g)<3.0$) stars observed in the same instrumental set-up, in order to be representative of those we are studying in this work. We then ran our complete pipeline on the retrieved spectra and obtained estimates of ${\rm T_{eff}}$, ${\rm \log(g)}$, [M/H], and [$\alpha$/Fe]. In the upper panels of \autoref{fig:ferre_ges_comp} we compare our set of parameters with those of GES. There is good agreement between both sets of estimates; comparisons follow linear trends reassuringly near to the 1:1 line with small dispersion and systematic scale differences. These plots show the ability of our pipeline to produce results of comparable quality as those of the GES. We also used these comparisons to estimate the small zero points in the stellar parameters compared to GES and these are shown in the figure. We applied these zero points to calibrate the parameters of our programme stars into the astrophysical scale of the GES survey.

\begin{figure}
    \centering
    \includegraphics[width=\figurefraction\columnwidth]{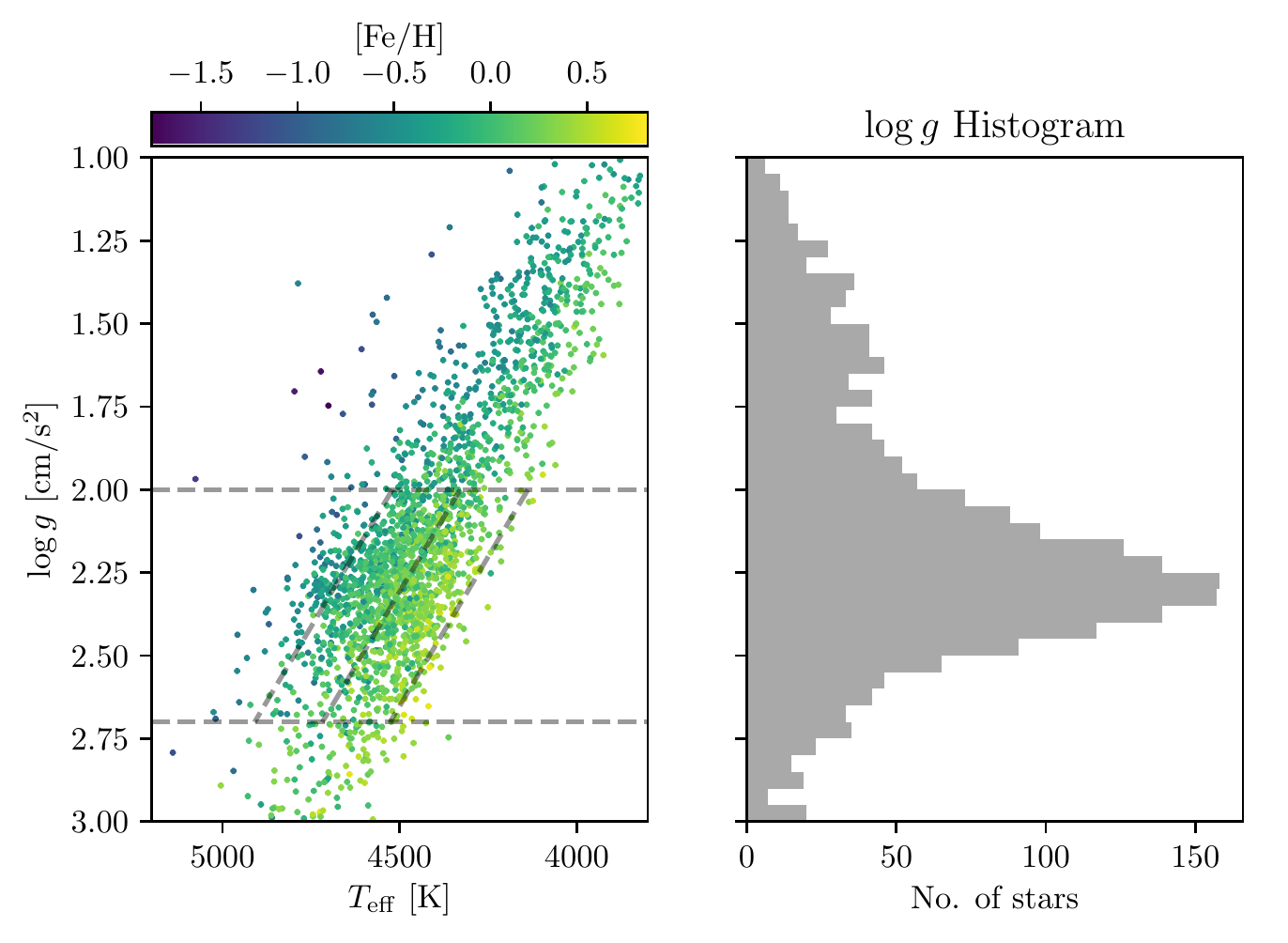}
    \caption{{\bf Left panel:} Distribution of calibrated ($\Teff$, $\logg$) colour coded by [Fe/H] for the entire sample. We also show the cuts used to define the sample of RCGs as dashed lines (see \autoref{eqn:cuts}); the diagonal cuts correspond from left to right to metallicities of $\FeH=-0.5,0.0,0.5$. {\bf Right panel:} Histogram of the surface gravity, $\log(g)$. The peak at $\log(g) \approx 2.3$ is due to RCGs and is clear and distinct.}
    \label{fig:stellarparameters}
\end{figure}

Once calibrated fundamental parameters were obtained for our programme stars, we adopted these parameters to estimate elemental abundances through line-by-line spectrum synthesis calculations using the BACCHUS code \citep{bacchus:ascl}. In the case of the \FeH iron abundances adopted in this letter, they were computed from the mean of individual estimates over five clean Fe I lines\footnote{Specifically, those at 8514.1, 8515.1, 8621.6, 8688.6 and 8846.7~\AA
.}. In the large lower panel of Fig.~\ref{fig:ferre_ges_comp}, we compare the results of this procedure applied on the $\sim600$ test stars from the GES. The comparison between our [Fe/H] estimates and those from the GES is linear and essentially unbiased. In consideration of that, we do not apply any zero point calibration to the [Fe/H] estimates of our programme stars.
In Fig.~\ref{fig:stellarparameters} we show the distribution of our programme stars in the Hertzsprung–Russell diagram, colour coded by iron abundance.
 
In a forthcoming paper (Rojas-Arriagada et al. in prep.) we will measure further elemental abundances beyond [Fe/H], and the complete sample and further details on our processing pipeline will be presented.

\section{Red clump giant sample selection}
\label{sec:rcgselection}

\begin{figure}
    \centering
    \includegraphics[width=\figurefraction\columnwidth]{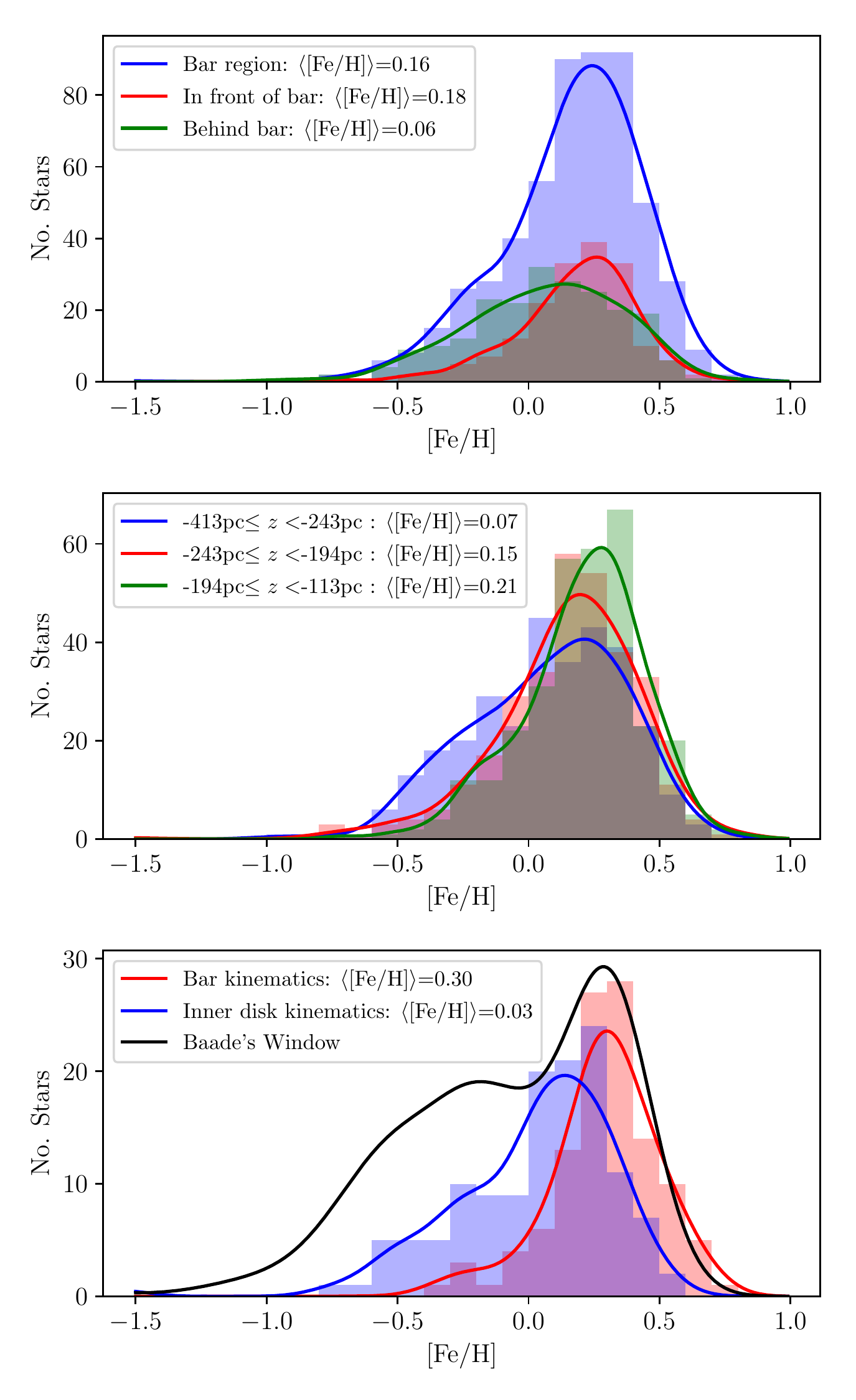}
    \caption{{\bf Upper panel:} Iron abundance distribution of stars within 0.5\kpc of the bar major axis (blue) compared to those in front and behind the bar (red and green, respectively). {\bf Middle panel:} Iron abundance distribution of stars in three vertical slices. The slices are chosen so each contains one-third of the sample. {\bf Lower panel:} Iron abundance distribution of stars spatially in the bar, separated kinematically into bar stars (red) and inner disc stars (blue; see \autoref{sec:barkin} for details). Although these stars are in the same region of space, the stars on bar orbits are on average more metal rich than their inner disc counterparts. In black we show the iron abundance distribution of Baade's window measured using the same instrument set-up \citep{Schultheis:17}. The highest metallicities in the bar are similar to those in the bulge, but there is a much smaller fraction of metal poor stars.}
    \label{fig:chem1d}
\end{figure}

In this paper we focus our analysis on RCGs. We use RCGs because first, they are standard candles with distances accurate to $\approx 10\%$ \citep{Bovy:14} and second, they form a very homogeneous sample, which makes differential analysis such as whether the bar is more metal rich than the disc, more straightforward. Therefore,
from the sample of 2456 stars we select a clean sample of RCGs by making very similar cuts in the (\Teff, \logg) plane to those described by \citet{Bovy:14} as follows:
\begin{align} 
& 2.0 < \logg< 2.7 \nonumber \\
&\logg \leq 0.0018 (\Teff - \Teffref)) + 2.5 \label{eqn:cuts} \\
& \Teff < 5000\,{\rm K,} \nonumber
\end{align}
where $\Teffref = -382.5\FeH + 4607\,{\rm K}$. The first of these cuts isolates clump stars in the \logg histogram (\autoref{fig:stellarparameters}), while the second removes first ascent giants. We also remove stars a small number of sample stars flagged as having problems with the NIR photometry or spectroscopic fitting.

We cross match this sample of 944 stars to Gaia DR2 in which every sample star has a companion within 0.35" (\autoref{fig:findingchart}). We remove the 6 clear foreground stars whose parallax places them closer than 3\kpc at 2$\sigma$,~\ie~those with $\varpi - 2\sigma(\varpi) > 1/3\kpc$.

We do not use the Gaia parallaxes beyond this foreground cut because they provide limited information at the distance of the bar for our sample. Instead we compute distances to our stars using the distance modulus 
\begin{equation}
\mu_K = \K - \overbrace{\frac{A_\K}{E(J-\K)} \underbrace{\left[ (J - \K) - (J-\K)_{\rm RC} \right]}_{\mbox{reddening}}}^{\mbox{extinction correction}}- M_{\K,{\rm RC}} ~,\label{eq:mujhk}
\end{equation}
where we adopt $M_{\K,{\rm RC}}=-1.61$ \citep{Hawkins:2017ko}, $(J-\K)_{\rm RC}=0.68$ and $\frac{A_\K}{E(J-\K)}=0.528$ \citep{gonzalez:11b}. An alternative approach would be to use spectro-photometric distances, however for RCGs the population corrections predicted by isochrones are highly uncertain \citep{Girardi:16,Chan:19}. Fortunately, since RCGs are  standard candles, these are at the $\sim0.05-0.1$ level and so we adopt the more straightforward approach of \autoref{eq:mujhk}. We also checked that we find similar results using distances computed using spectro-photometric distances derived using the method of \citet{RojasArriagada:17}. We also checked that selecting RCGs from APOGEE DR14 and applying \autoref{eq:mujhk} gives distances that are consistent with the spectro-photometric distances computed by \citet{Queiroz:18} with a mean absolute deviation of less than 9\%.
Our final sample comprises 938 RCGs stars with full 6D kinematic information together with spectroscopic iron abundance measurements.

\section{Chemistry of the long bar of the Milky Way}
\label{sec:barchem}
\subsection{Spatial selection of bar stars}

To investigate how the chemistry of the bar compares to that of the inner disc we use the 3D position of our stars to divide the sample into three parts: 
\begin{inparaenum}[(i)]
        \item Those in the bar region, defined to be within $\pm0.5\kpc$ of the major axis of the bar. In making this selection we assume $R_0=8.2\kpc$ \citep{Gravity:19} and that the bar major axis lies at $27\deg$ \citep{Wegg:13} to the Sun-galactic centre line of sight. 
        \item Those in front of this bar region, but still more than $3\kpc$ distant from the Sun.
        \item Those behind this bar region, but still less than $10\kpc$ distant from the Sun.
\end{inparaenum}
For reference, at a longitude of $l\sim18\deg$ typical for our sample, these correspond to (i) bar stars with distance $4.6\kpc < d < 5.8\kpc$, (ii) stars in front of the bar with $3\kpc < d < 4.6\kpc$, and (iii) stars behind the bar with $5.8\kpc < d < 10\kpc$.

In the upper panel of \autoref{fig:chem1d} we show the iron-abundance distribution of each of these sub-samples. We find that the metallicity distribution function (MDF) of the stars in the bar and in front of the bar are very similar, while those behind the bar are more metal poor. Quantitatively the mean \FeH of the bar stars is $0.16\pm0.01\,{\rm dex}$ compared to $0.18\pm0.02\,{\rm dex}$ for those in front of the bar and $0.06\pm0.02\,{\rm dex}$ for those behind.

However, we also find a very strong vertical metallicity gradient. We show the distribution of \FeH in three vertical slices in \autoref{fig:chem1d}. The population of stars changes as a function of distance to the galactic plane; a higher fraction of more metal poor stars is found in the slice furthest from the galactic plane. Fitting a linear relation between \FeH and distance from the galactic plane across the entire sample, the vertical gradient is $-1.1\pm0.2\,{\rm dex}\,{\rm kpc}^{-1}$.

Because we use pencil beam fields just away from the galactic plane, our more distant stars are also on average further from the galactic plane. We must therefore account for this vertical gradient to compare the [Fe/H] distribution in the bar to the surrounding inner disc. Using the measured vertical gradient of $-1.1\,{\rm dex}\,{\rm kpc}^{-1}$ to correct our entire sample to $200\,{\rm pc}$ from the galactic plane, we find the mean \FeH of the stars in the bar region is $0.18\pm0.01\,{\rm dex}$ compared to $0.15\pm0.02\,{\rm dex}$ for those in front of the bar and $0.15\pm0.02\,{\rm dex}$ for those behind.

We conclude that, at the same distance from the galactic plane, the bar region in our sample is slightly more metal rich than the surrounding disc. We note that this conclusion is the opposite from that found by \citet{Bovy:19}, a disagreement we discuss further in \autoref{sec:disc}.

\subsection{Kinematic selection of bar stars}
\label{sec:barkin}

\begin{figure}
    \centering
    \includegraphics[width=\columnwidth]{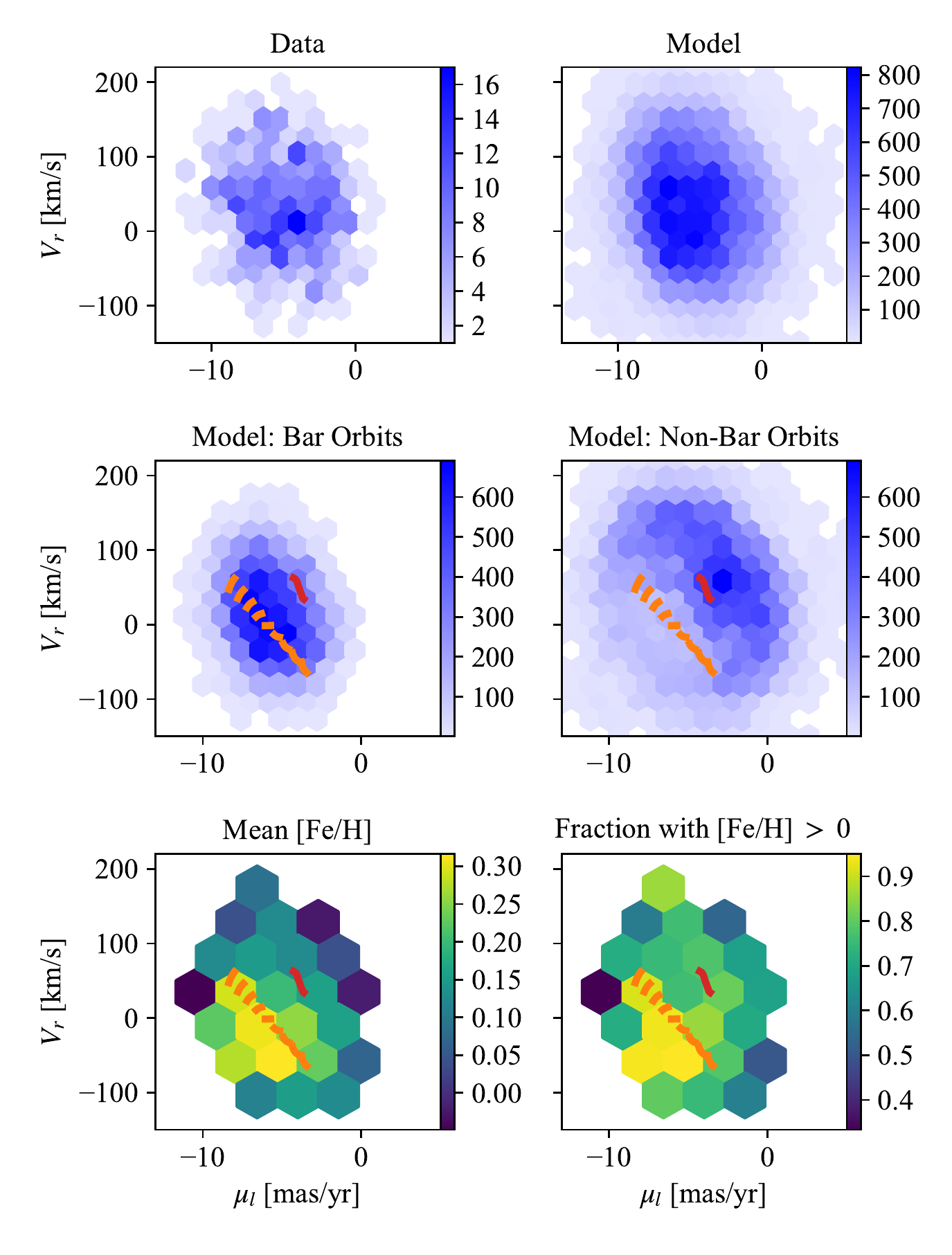}
    \caption{{\bf Upper panels:} Kinematics of sample stars (left) and model stars (right) within 0.5\kpc of the bar major axis. {\bf Middle panels:} Model stars separated into whether they are on bar orbits (see text for details). {\bf Lower panels:} The mean metallicity (left) and fraction of  super solar metallicity stars (right). In the middle and lower panels we also plot the model circular velocity at these galactocentric distances (red) and orbits with zero angular momentum in a frame rotating at $\Omega=35\kms\kpc^{-1}$ (orange). These zero angular momentum orbits correspond to an pencil thin bar and have galactocentric velocity -80\kms to 80\kms in steps of 20\kms. In the model, bar orbits have kinematics near these low angular momentum orbits, and the stars in this region of kinematic space are preferentially metal rich.}
    \label{fig:kin}
\end{figure}

In this section we present the 2D kinematics of bar stars and make a kinematic selection of stars whose orbits indicate they follow the bar. To guide our selection and methods we use a barred made-to-measure Milky Way model of the Galaxy, which was fitted to a range of data on the inner Galaxy by \citet[][hereafter \citetalias{Portail:17}]{Portail:17}. We note however, that this model was fit only to star count data in the long bar (the region studied in this work) because of the lack of kinematic data at that time. We choose to show a model with a pattern speed of $\Omega=35\kms\kpc$ because its kinematics agrees qualitatively slightly better than the fiducial model ($\Omega=40\kms\kpc$), however our conclusions would remain unchanged with any model from \citetalias{Portail:17}. To compare this model to our data we use the Galaxia code \citep{Sharma:11} to generate mock catalogues of stars generated from the Padova isochrones with a 10\,Gyr old population. For our fiducial MDF we use a Gaussian with $\langle\FeH\rangle=0$, $\sigma(\FeH) = 0.3$. To these mock catalogues we apply our selection criteria, both spatially and in the ($T_{\rm eff}$, \logg) plane, to generate a mock sample of RCG stars. We also compute distances to this mock catalogue using \autoref{eq:mujhk} in the same manner as the real sample. In the mock sample the mean absolute deviation of this RCG distance to the true distance is 6.1\%. To replicate the treatment of the real sample we use these RCG distances when using the mock catalogue in what follows.

In the upper left panel of \autoref{fig:kin} we show the kinematics of the stars selected to be in the bar region in the longitudinal proper motion versus radial velocity plane (i.e. $\mu_l,V_r$ plane). This 2D plane encodes similar information as the well-known $(U,V)$ plane locally. 
We compare this to the mock sample finding qualitatively very good agreement. The sample is centred at $\mu_l\approx-6\,{\rm mas}\,{\rm yr}^{-1}$ similar to the $\mu_l=-6.24\,{\rm mas}\,{\rm yr}^{-1}$ of Sgr A* \citep{Reid:2004bi}. This is largely due to the motion of the Sun with respect to the galactic centre and it is reassuring for the purity of our sample that there is a complete lack of stars at $\mu_l>0\,{\rm mas}\,{\rm yr}^{-1}$ where nearby foreground contaminants would lie.

Using our mock sample of stars we can investigate the kinematic signatures of the bar in this 2D kinematic plane. We integrate the model stars and define long bar stars as those which spend more than 40\% of their time in the long bar region; that is within $\pm0.5\kpc$ of the bar major axis and more than 2\kpc from the galactic centre. We also used the orbital frequency based definition from \citet{Portail:2015dj} (specifically $f_r/f_x\in2\pm0.1$) finding very similar results. However we use the orbital time definition because it both better selects particles which were human-classified to be on bar orbits, and better photometrically separates bar stars from inner disc stars in face-on projections.  

We split the model based on this orbital selection of bar stars in the middle panels of \autoref{fig:kin}. In these panels we also plot the position of the circular velocity for our $l=18\deg$ longitudinal field (the red line), together with the orbits of stars travelling directly along the major axis of the bar, i.e. with zero angular momentum in the co-rotating frame (orange lines). These pencil thin bar orbits cover a galactocentric radial velocity range -80\kms to 80\kms in steps of 20\kms. As expected, the model bar orbits lie in the region of these pencil thin orbits with a scatter due to the width of the bar and its underlying orbits \citep{Valluri:16}.

In the lower panel of \autoref{fig:kin} we plot the iron abundance of our sample in the same plane. In the region of kinematic space where the bar stars lie, both the mean metallicity and the fraction of stars with super-solar abundance ($\FeH>0$) appear significantly higher.

To quantify these differences we use the $\mu_l,V_r$ plane shown in \autoref{fig:kin} to assign each of our sample stars a probability, $p_{\rm bar}$, that it is on a bar following orbit. We do this by for each sample star computing the fraction of bar following orbits among the 100 nearest neighbour mock sample stars in this plane. This provides a sample of stars with (\FeH,$p_{\rm bar}$). We use this to plot in the lower panel of \autoref{fig:chem1d} the MDF of stars with bar orbits, $P(\FeH | p_{\rm bar} = 1)$, and inner disc orbits, $P(\FeH | p_{\rm bar} = 0)$. We compute this using a conditional kernel density estimate (KDE) $P(\FeH | p_{\rm bar}) = P(\FeH, p_{\rm bar}) / P( p_{\rm bar} ),$ where both probability distributions on the right-hand side are estimated using a Gaussian KDE. We also plot histograms of the stars with the highest probability to be bar stars, $p_{\rm bar}>0.8,$ and the lowest, $p_{\rm bar}<0.2$. These are very similar to the KDE estimated bar and inner disc MDFs.

Using this kinematic selection of bar stars we find that the bar is more metal rich than the inner disc with a mean iron abundance of $\langle \FeH \rangle = 0.30$ compared to 0.03 for the kinematically selected inner disc stars.

\section{Discussion and conclusions}
\label{sec:disc}

We find that the bar of the Milky Way outside the B/P bulge (the so-called long bar) is more metal rich than the inner disc. We find this both kinematically and spatially selecting bar stars. The kinematic sample of bar stars is extremely metal rich with $\langle \FeH \rangle = 0.30$. This MDF is comparable to the metal rich part of the MDF measured in the bulge, but unlike in the bulge, there are very few metal poor stars; only 10\% of our kinematically selected bar stars have sub-solar iron abundance.

 In contrast, the sample of stars selected spatially to be within the bar are only slightly more metal rich (with $\langle \FeH \rangle = 0.18$ at $z=200$pc) compared to foreground and background stars (which had $\langle \FeH \rangle = 0.15$).  Spatially the bar and inner disc samples have a more similar mean metallicity than the kinematic selection for two reasons. First, the spatially selected samples of bar stars and inner disc stars will both contain a mixture of stars with bar orbits and inner disc orbits. Second, the distance errors are $\sim$10\%, which is larger than the kinematic errors and will tend to scatter stars between the samples, reducing the contrast.
 
 To test the consistency of our kinematic and spatial results, we therefore constructed a mock survey where the bar stars had a Gaussian MDF with $\langle\FeH\rangle=0.35$, $\sigma(\FeH) = 0.2$ at $z=200$pc and the inner disc stars had a Gaussian MDF with $\langle\FeH\rangle=0.07$, $\sigma(\FeH) = 0.3$ at $z=200$pc. Both populations were given a vertical gradient of $d\langle \FeH \rangle/d|z| = -1{\rm dex}/\kpc$ and the sample was constructed using Galaxia to sample the P17 models as described in \autoref{sec:barkin}. After applying our selection function, this mock sample was analysed just as the real sample. We find that the kinematically selected bar stars in the mock sample have $\langle \FeH \rangle = 0.32$ while the inner disc has $\langle \FeH \rangle = 0.02$; these are very similar values to the real sample, which has $0.30$ and $0.03,$ respectively. However, the spatially selected mock sample has a much more similar mean metallicity: $0.16$ in the bar region at $z=200$pc compared to $0.12$ in front of the bar and $0.12$ behind. This simple model shows that our sample indicates that the bar is more metal rich than the inner disc in a consistent manner both when kinematically and spatially selecting bar stars.
 
A metal rich bar has several interesting interpretations: (i) If these stars were in place at the time of bar formation, then stars which make up the bar would preferentially sample the kinematically coolest populations \citep{Debattista:17,Fragkoudi:17}. Assuming that the inner disc at bar formation had a metallicity-age-dispersion relation analogous to that seen locally now, where young stars are preferentially metal rich with low dispersion and scale height, it would then be natural that the bar would be more metal rich than the inner disc.

However this picture is influenced by two further effects: (ii) bars in simulations grow by capturing stars from the disc over their life, particularly in the long bar region \citep{Aumer:15}. These captured stars are likely to be kinematically cool stars with  guiding radii just outside the bar, a region whose stars are known to be metal rich \citep[e.g.][]{Hayden:15}; and (iii) star formation in the bar region preferentially occurs in gas clouds that trace the bar together with associated dust lanes \citep{Martin:97}. These recently formed stars are therefore preferentially be bar following and are also likely be metal rich.

Understanding the relative importance of these effects will require further data. Ages, or age indicators, in particular would help clarify if the metal rich stars formed early, together with the inner disc as expected by (i), or instead have an extended star formation history as expected by (ii) or (iii).

It is noteworthy that a bar which extends outside a B/P bulge and is more metal rich than the surrounding inner disc is also seen in our neighbouring barred galaxy, M31 \citep{saglia:18}. This situation is also common in other nearby barred spirals \citep{Gadotti:18}. 

We note that our results appear to be in disagreement with the recently submitted paper by \citet{Bovy:19} who finds that the bar is more metal poor than the surrounding disc. The reason for this apparent disagreement is unclear, but we emphasise the homogeneity of our RCG sample.

\begin{acknowledgements}
      CW acknowledges funding from the European Union's Horizon 2020 research and innovation program under the Marie Sk\l{}odowska-Curie grant agreement No 798384.\\
      
     Based on observations collected at the European Southern Observatory under ESO programme 099.B-0532(A).\\
      
       This  work presents  results  from  the  European  Space  Agency  (ESA) space  mission Gaia.  Gaia  data  are  being  processed by  the Gaia  Data  Processing  and  Analysis  Consortium  (DPAC). Funding for the DPAC is provided by national institutions, in particular the institutions participating in the Gaia MultiLateral Agreement (MLA).\end{acknowledgements}

%
%

\end{document}